# Record-high photocatalytic rate at single metallic atom oxide

Cong Wang[1], Ang Li[1], Chong Li[2], Shengbai Zhang[3], Hui Li[1], Liming Hu[4], Yibo Feng[1], Kaiwen Wang[1], Zhu Zhu[1], Ruiwen Shao[5], Yanhui Chen[1], Peng Gao[5], Shengcheng Mao[1], Jun Huang[1,6], Ze Zhang[7] and Xiaodong Han[1*]

[1]Institute of Microstructure and Property of Advanced Materials, Beijing Key Lab of Microstructure and Property of Advanced Materials, Beijing University of Technology, Beijing, 100124, China

[2]International Laboratory for Quantum Functional Materials of Henan, School of Physics and Engineering, Zhengzhou University, Zhengzhou, 450001, China

[3]Department of Physics, Applied Physics & Astronomy, Rensselaer Polytechnic Institute, Troy, NY 12180, USA

[4]Department of Biological and Chemical Engineering, Beijing University of Technology, Beijing, 100124, China

[5]Electron Microscopy Laboratory, and International Center for Quantum Materials, School of Physics, Peking University, Beijing 100871, China

[6]The University of Sydney Nano Institute, Sydney, New South Wales 2006, Australia

[7]Department of Material Science, Zhejiang University, Hangzhou, 310008, China

[*]Corresponding author: X. D. H. (xdhan@bjut.edu.cn)

Metal oxides, as one of the mostly abundant, low-cost and widely utilized materials, have been extensively investigated and applied in environmental remediation and protection, energy conversion and storage etc[1-7]. Most of these diverse applications are results of a large diversity of the electronic states of metal oxides. Noticeably, however, numerous metal oxides have obstacles for applications in catalysis because of low density of active sites in these materials. Size reduction of oxide catalyst is a widely-adopted strategy to improve the active site density[8,9]. Here, we demonstrate the fabrication of single tungsten atom oxide (STAO) in which the oxide's size reaches its minimum. However, the catalytic mechanism in this STAO is determined by a quasi-atom physics which is fundamentally distinct from the traditional size effect, and also is in contrast to the standard condensed matter physics. STAO results in a record-high and stable sunlight photocatalytic degradation rate of $0.24$ $s^{-1}$, which exceeds those of available photocatalysts by approximately two orders of magnitude. The photocatalytic process is enabled by a quasi-atom physical mechanism, in which, an electron in the spin-up channel is excited from HOMO to LUMO+1 state (both are largely tungsten atomic d orbitals), which can only occur in STAO with $W^{5+}$. The fabrication of STAO and the discovered unique quasi-atom physics lays a new ground for achieving novel physical and chemical properties using various single metallic atom oxides.

Pushing the size and dimension reduction of oxide to its limit, i.e., the single metallic atom form of metal oxide, can ultimately enhance the density of active catalytic sites. More importantly, a rich tunability of oxygen coordination may fundamentally alter the catalytic mechanisms. For example, the catalytic behaviour will transition from the properties of a solid to those of a quasi-atom, namely those of metallic atom and its coordinated oxygen atoms. However, single metallic atom oxide has yet to appear, although recent achievements have been made in single metallic atom catalysts on a support[10-21]. In this paper, we report highly stable single tungsten atom oxide (STAO) photocatalyst, which is uniformly dispersed in aqueous solution as a byproduct of our co-ordinated route, where polyethylene oxide, as a support, was specifically designed to anchor STAO. Moreover, the interaction between the coordinated anchoring support and tungsten atom would not only form single metallic atom oxide with mono-dispersity but also affect the photoelectronic states of STAO. To illustrate the strategy of STAO with respect to the traditional semiconductor catalyst and single metallic atom catalyst on substrates, Fig. 1a shows an example of conventional photocatalytic oxide: single crystal tungsten oxide and its electronic state of conduction and valence band. Fig. 1b illustrates that STAO exhibits a tungsten highest occupied molecule orbital (HOMO) state near Fermi level. The intra-atomic d-orbitals allow an electron transition in the photo-excited STAO. Fig. 1d shows a single Pt atom on the $TiO_2$ oxide which derived from the pure Pt metals of Fig. 1c. Comparing to Fig. 1a, c and d, note that STAO in Fig. 1b introduces a new quasi-atom mechanism in which the spin-up and spin-down electrons in d-orbitals can transit within the oxidized single tungsten atom, which is fundamentally different from the inter-d band electron transfer in single metallic atom materials (Fig. 1d) and the inter-atomic $O_{2p}$-$W_{5d}$ electron transfer in conventional $WO_3$ crystal (Fig. 1a). Expectedly, the excited states of intra-d-orbital electrons from the quasi-atom mechanisms combining with the maximized density of active sites from STAO can yield an ultra-high catalytic rate under illumination.

Our synthesized STAO (see Supplementary Method) is shown in Fig. 2a, in which the mono dispersed nature was characterized by scanning transmission electron microscopy (STEM). The single tungsten atoms can be observed as bright spots in the high-angle annular dark field (HAADF) images (Fig. 2b). X-ray photoelectron spectroscopy (XPS) and electron energy loss spectroscopy (EELS) are employed to reveal the chemical states of tungsten and oxygen. The binding energy between W and O is determined by the XPS spectra (Fig. 2c). Different from the spectrum of pure $W^{6+}$ (Extended Data Fig. 1), the spectrum obtained from STAOs can be divided into two doublets. The main peaks are the $W^{6+}$ doublet at 35.8 eV ($W_{4f}7/2$) and 37.8 eV ($W_{4f}5/2$), while the $W^{5+}$ doublet appears as shoulders at 34.7 eV ($W_{4f}7/2$) and 36.7 eV ($W_{4f}5/2$)[22]. The chemical state of oxygen is examined by electron energy loss spectroscopy (EELS), as presented in Fig. 2d. As a comparison, the spectrum from commercially available $WO_3$ particle as a reference sample, is shown as the black solid line. Compared with the EELS spectrum of the $WO_3$ particle reference, the STAO spectrum shows that P1, P2 and P4 diminish. The disappearance of the P1 and P2 may be attributed to the fact that in $WO_3$ particles, the transition is from $O_{2p}$ to $W_{5d}$[23], which disappears in STAOs. On the other hand, the disappearance of P4, which is a characteristic of the second nearest neighbor W-O-W peak[24], is consistent with mono dispersing STAO without nearest neighboring W atoms.

The STAO exhibits superior catalytic performances for the photocatalytic degradation of dyes (methyl orange, methyl red and dimethyl yellow, shown in Extended Data Fig. 2). The photocatalytic rate under simulated sunlight (98.5 mw/cm$^2$ detailed characterization in Supplementary Method) of 0.24 s$^{-1}$ exceeds those of available nano-sized photocatalysts by two orders of magnitude, as summarized in Fig. 3a and Extended Data Table 1. The performance of STAO can be directly visualized in Extended

Data Fig. 3 and Movies 1, 2. It is demonstrated in Fig. 3b that the 365nm-appearent quantum yield (AQY, 1.84 mw/cm²) of STAO is also at least two orders of magnitude larger than that of commercial TiO₂ nanoparticle (P25) at the same density (detailed spectra are shown in Extended Data Fig. 4.). STAO has significantly less light scattering, and the interactions among STAOs are much weaker than those among nanoparticles when the density of the catalyst comes to a saturation. Consequently, P25 tends to reach its efficiency maximum at a low saturation density (Extended Data Fig. 5), while STAO's efficiency increases linearly as its density increases. When it comes to 0.2 mol/L, the AQY of STAO of ~64.82% is achieved and external quantum efficiency is shown as Extended data Fig. 6. To verify the stability and robustness of the catalyst, 1st to 5th photocatalytic degradation (see Supplementary Method) results of STAO showed no significant decrease in efficiency (Extended data Fig. 7). Furthermore, the high resolution TEM images of the durable photocatalyst indicate that the single metallic atom oxide has been kept as mono-dispersity after 5 times repeated degradation performances (Extended Data Fig. 8a). Extended Data Fig. 8b, c show that the photocatalyst with 5 times-recycle run has no observable change in chemical valence state and the radical signal in electron spin resonance (ESR). Additionally, the photocatalyst are also robust on various supports (Tween, PEO-PPO-PEO and TiO₂), and all of them exhibit high and stable photocatalytic efficiencies (see Extended Data Fig. 9a-c and Supplementary method). Extended Data Fig. 9d-f show that the monodispersed metal oxides on the supports are intact after recycles. The ESR and XPS data in Extended Data Fig. 9g, h further indicate that the physical properties of the STAOs on other supports also have not changed.

Our experiments raise three fundamental questions: (1) the microscopic origin of W⁵⁺ in relation to the coordinated anchoring of STAOs on polyethylene oxide, (2) the physical nature of HOMO-LUMO (lowest un-occupied molecular orbitals) states of STAO in relation to the oxidation states of tungsten, and (3) the electron transfer mechanism of W⁵⁺ in the STAO limit.

The large amount of STAOs in our experiments suggests that their formation is energetically favored via a coordination route with polyethylene oxide. STAOs can vary in their tungsten valency between 5+ and 6+ by a different hydrogen attachment, e.g., $W^{5+}$ in $WO_4H_3$ and $W^{6+}$ in $WO_4H_2$. The first-principles total-energy calculations for the coordination suggests that the formation energy (see Supplementary Table S2) for $W^{6+}$ is -0.68 eV, whereas that for $W^{5+}$ is -0.69 eV. Both configurations prefer the distorted octahedron geometry as a result of strong binding between STAOs and polyethylene oxide, as well as all other supports. In our experiment, the STAO-polyethylene oxide complexes are often present in a solution which facilitates the conversion between $W^{5+}$ and $W^{6+}$. This process is demonstrated in Fig. 4a where a $W^{6+}$ complex reacts with one hydroxyl to form a $W^{5+}$ complex plus 1/2 $H_2O_2$ through proton coupled electron transfer. We have performed ESR measurements in the dark (see Extended Data Fig. 10a-c), which confirms the presence of dimethyl pyridine N-oxide-OH radicals. This is a direct evidence that $H_2O_2$ is present in $W^{5+}$ complex. On the other hand, the Extended Data Fig. 11, 12 shows that, the catalytic efficiency increases considerably when the pH decreases from ~7 to 1, which signals an increase in the concentration of the $W^{5+}$ complexes.

In light of the different hydrogen coordination, one can expect noticeable changes in the absorption spectrum. Our calculation yields a gap of 4.96 and 3.03 eV for $W^{6+}$ and $W^{5+}$ (Fig. 4b), respectively, which matches the measured absorption peak at 5.11 and 3.02 eV (see Extended Data Fig. 13). The marked difference in the HOMO-LUMO gaps between $W^{5+}$ and $W^{6+}$ suggests that one could determine the role of $W^{5+}$ and $W^{6+}$ in the dye degradation reaction by tuning the light wavelength. Photocatalytic degradation experiments were carried out with a 230nm (5.39 eV) laser to preferentially excite $W^{6+}$ and, separately, with a 380nm (3.26 eV) laser to preferentially excite $W^{5+}$. A similarly excellent efficiency obtained at 380nm light demonstrates that the STAO complexes with $W^{5+}$ play the dominant role in the reaction.

In order to understand why STAO complexes with $W^{5+}$ possess a high catalytic efficiency, we calculate the electronic orbitals for STAO complexes. The additional electron of $W^{5+}$ occupies a spin-up HOMO state in Fig. 4b, which is empty in $W^{6+}$, and results in two important consequences: (a) $W^{5+}$ is spin-polarized with 1μB. The presence of spin dictates that the singly-occupied HOMO states spin split. (b) The $W^{5+}$ HOMO state is a tungsten d state, oppose to the oxygen p state for $W^{6+}$. Our analysis shows that the HOMO state of $W^{5+}$ is made of tungsten $d_{xz}$ and $d_{yz}$ orbitals, the LUMO state is made of tungsten $d_{xy}$ and $d_{x2-y2}$ orbitals, while the LUMO+1 state is made of a tungsten $d_{yz*}$ orbital (see yellow clouds in Fig. 4b). While optical transitions between different spin channels are forbidden, within the same spin channel, most of the low-energy transitions are also forbidden due to local d-orbital symmetry of the wavefunctions, except for the transitions from HOMO to LUMO (very weak) and from HOMO to LUMO+1 (strong). Note that due to the strong localization of d orbitals in atomic-like STAOs, here one cannot neglect excitonic effect when calculating optical transitions. As it turns out, the effect is large, e.g., 4.53 eV for the HOMO to LUMO+1 transition to result in an optical gap of 3.03 eV. Similarly, when calculating the transition for the $W^{6+}$ states, excitonic effect is also important and the minimum transition energy gap is 4.96 eV and is between HOMO-6 (an oxygen p state) and LUMO. The qualitative difference between $W^{5+}$ and $W^{6+}$ corroborates with the EELS results in Fig. 2d, showing that the P1 and P2 peaks for bulk $WO_3$ ($W^{6+}$) are absent in STAO ($W^{5+}$). When optical excitation creates electron-hole pairs in STAO with $W^{5+}$ (Fig. 4c), the photogenerated holes split H-OH to result in the formation of hydroxyl radicals[25,26]. The photogenerated electrons, on the other hand, would cause a selective cleavage of the C-N bonds in the azo-benzenesulfonic group of the methyl orange molecules[27,28]. The resulting hydroxyl radicals and benzenesulfonic radicals combine to form p-hydroxybenzenesulfonic acid (m/z=172, HPLC-MS, negative mode). Should this happen, *N,N*-dimethylaniline (m/z=122, HPLC-MS, positive mode) should also present in the final products. Experimental results in Extended Data Figs. 14, 15 showed that both *N,N*-dimethylaniline and p-hydroxy benzenesulfonic acid are detected as the dominant reaction products. The above catalytic route is not only in full agreement with experiments but also supported by first-principles calculations. Fig. 4d shows that, among the various possible cleavage (i)-(iv), pathway (i) is energetically favoured.

In summary, a record-high catalytic rate of 0.24 $s^{-1}$ under sunlight and a 64.82% AQY at 365nm are approached with stability, which both exceed those of available photocatalysts by at least two orders of magnitude. The record-high efficient photocatalytic degradation is mainly attributed to high density of active catalytic sites and the novel quantum physics in STAO. The properties of the STAO critically depend on the coordination with oxygen, as well as different from bulk oxide for it replaces bulk band edge p-d transition by pseudo-atomic d-d transition. Also important is the simple method to anchor coordinated single metallic atom oxides on supports. The methodology for the anchoring of STAOs is readily adopted for a wide-range transition metal oxide (such as Mo, Nb, Zr shown in Extended Data Fig. 16), and anticipated to chalcogenides and halides with outstanding properties for catalysis, energy conversion, information storage, environment protection, chiral optics and quantum information science.

Acknowledgements

Work in China was supported by the Key Project of CNSF (11234011), the CNSF (11374030, 11327901, 11127404), the Beijing Nova Program (Z151100003150142), China, the Project of Construction of Innovative Teams (IDHT20140504). S. B. Z. was supported by US National Science Foundation under Grant No. DMREF-1627028. We thank Prof. Z. M. Guo, Prof. Y. Wang, Prof. P. F. Wang, Prof. Z. Sun, Prof. Z. G. Yan, A. D. Wang for providing measurement support, and we also thank Prof. S. X. Ouyang, Prof. L. H. Wang, Prof. K. Zheng, Dr. S. Y. Chen for discussions.

Author Contributions: C. W. and X. D. H initiated, conceived and guided the research. S. B. Z. led the DFT calculations and theoretical analysis. C. L. performed the calculations. C. W., H. L. and Y. B. F. designed and synthesized the molecular precursors, A. L., Y. H. C., R. W. S., P. G., S. C. M. Z. Zhang and X.D.H performed TEM characterization and analysis by finding the single metallic atomic oxides. C. W., L. M. H., K. W. W., Z. Z. and J. H. performed catalytic experiments and analysis. C. W., A. L., C. L., S. B. Z, and X. D. H. wrote the manuscript. All authors contributed to the scientific discussion.

Competing interests: The authors declare no competing interests.

Correspondence and requests for materials should be addressed to X.D.H

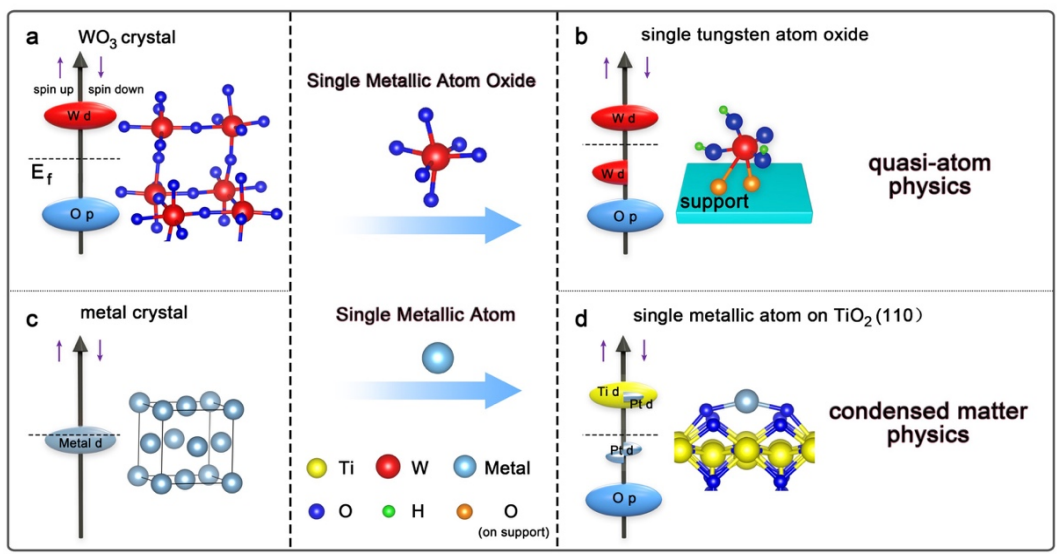

**Fig. 1. Single atomization strategy of metal oxide.** Electronic properties and its structure of $WO_3$ crystal (a), single tungsten atom oxide (b), noble metal crystal (c) and single Pt atom on $TiO_2(110)$ (d). The single metallic atom oxide approach is choosing appropriate support. The electronic property of STAO is distinct, compared with those of the metal oxide crystal and single metallic atom on support.

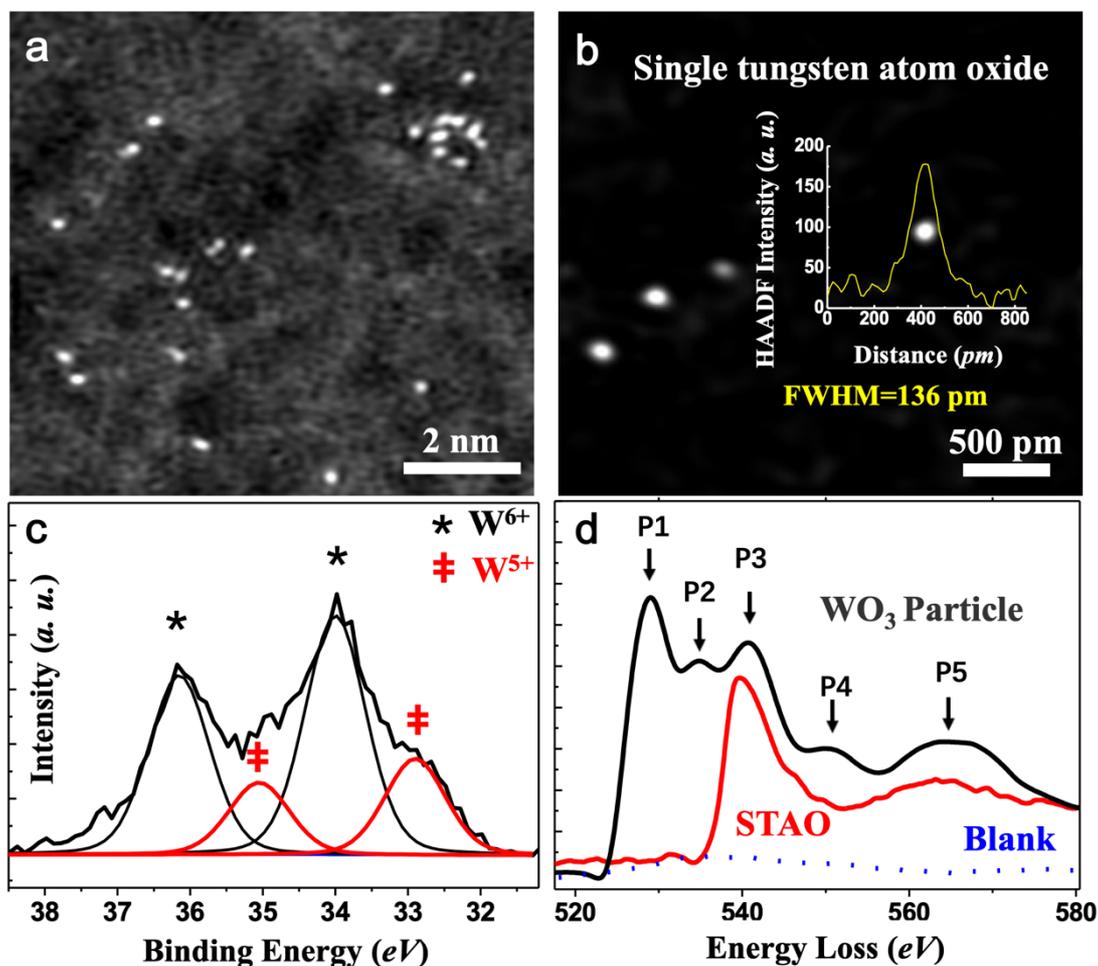

**Fig. 2. Characterizations and chemical analysis of STAO. a,** A Cs-corrected STEM-HADDF image shows monodispersed single tungsten atoms. **b,** A Cs-corrected STEM-HADDF image at a larger magnification. Inset shows the brightness of the full width half maximum (FWHM), which is consistent with the diameter of a tungsten atom. **c,** XPS indicates that tungsten in STAOs has two valence states of 5+ and 6+. Black line is $W^{6+}$ and red line is $W^{5+}$. **d,** EELS data of O K-edge of STAOs (red line). For a comparison, those of $WO_3$ particles (black line) and blank (blue line) are also shown.

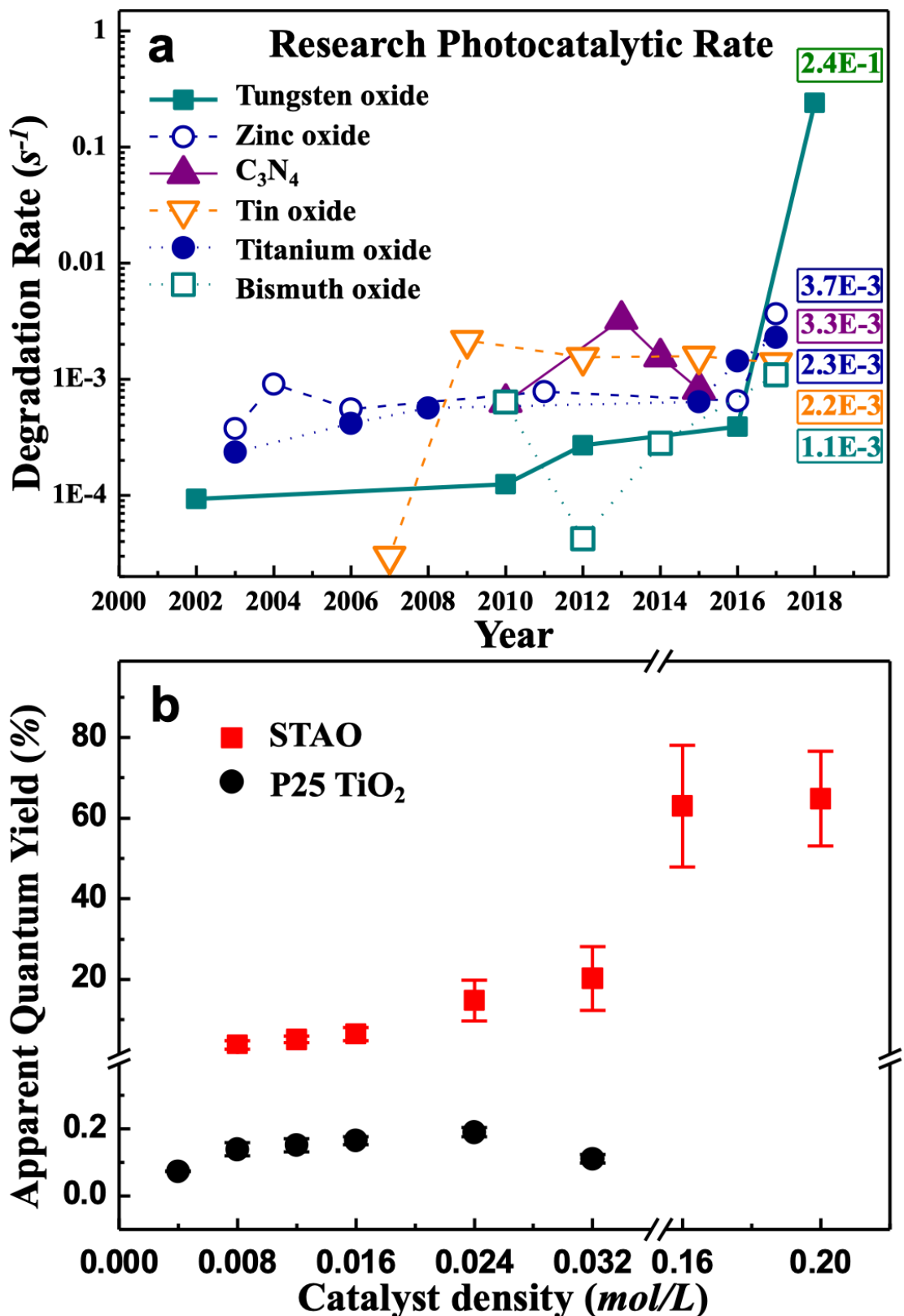

**Fig. 3. Photocatalytic degradation performance of STAO. a,** Photocatalytic rate of dye by available catalysts under sunlight (AM1.5G). **b,** AQY as a function of catalyst density. Note that AQYs of P25s are at least 100 times smaller than those of STAOs. Error bars represent standard deviation given by at least three independent measurements.

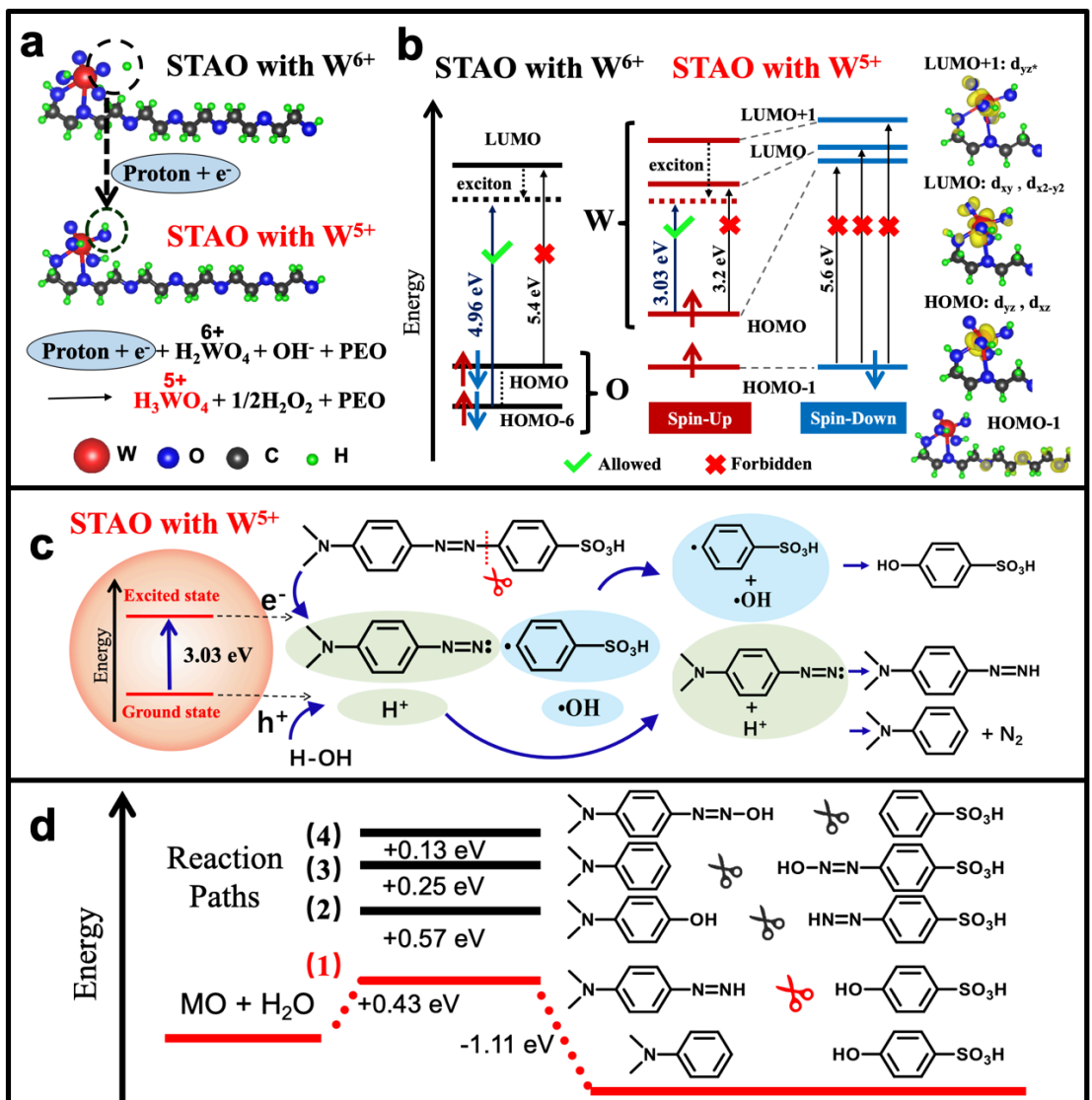

**Fig. 4. Mechanism for record-high rate at STAOs. a,** Schematic illustration of $W^{5+}$ formation in STAO. Red, blue, black and green balls are tungsten, oxygen, carbon, and hydrogen atoms, respectively. **b,** Occupations and transitions between states in $W^{6+}$ (left panel) and separately in $W^{5+}$ (middle panel). Arrows ↑ and ↓ denote spins. Dashed lines connect spin-up and down channels in the same band. The HOMO, LUMO and LUMO+1 bands for $W^{5+}$ have been decomposed (right panel) according to tungsten atomic orbitals. **c,** Photocatalytic degradation reaction: upon photoexcitation, the hole dissociates an $H_2O$ into $H^+$ and ·OH. The electron cleaves a C-N bond in the azo-benzenesulfonic group. The resulting $H^+$ then reacts with 4-(2λ²-diazenyl)-*N,N*-dimethyl-aniline (green) to form *N,N*-dimethylaniline, while the ·OH reacts with benzenesulfonic acid radical (blue) to form p-hydroxy benzenesulfonic acid. **d,** Four possible cleavage pathways (1)-(4) with calculated activation energies. It shows that the $N_2$ release via a proton transfer reaction [i.e., path (1) marked in red] is energetically favoured.